\documentclass[letterpaper]{article}

\usepackage{proceed2e}
\usepackage{times,cuted}
\usepackage{epsfig}
\usepackage{graphicx}
\usepackage{amsmath}
\usepackage{amssymb}

\usepackage[pagebackref=true,breaklinks=true,letterpaper=true,colorlinks,bookmarks=false]{hyperref}

\begin{document}

\title{Sentient Self-Organization: Minimal dynamics and circular causality}

          
\author{ {\bf Biswa Sengupta}\\
Dept. of Bioengineering, Imperial College London\\
Dept. of Engineering, University of Cambridge\\
Cortexica Vision Systems Ltd.\\
{\tt\small b.sengupta@imperial.ac.uk}
\And
{\bf Karl Friston}\\
Wellcome Trust Centre for Neuroimaging\\
University College London\\
{\tt\small k.friston@ucl.ac.uk}
}

\maketitle

\begin{abstract}
Theoretical arguments and empirical evidence in neuroscience suggests that organisms represent or model their environment by minimizing a variational free-energy bound on the surprise associated with sensory signals from the environment. In this paper, we study phase transitions in coupled dissipative dynamical systems (complex Ginzburg-Landau equations) under a variety of coupling conditions to model the exchange of a system (agent) with its environment. We show that arbitrary coupling between sensory signals and the internal state of a system -- or those between its action and external (environmental) states -- do not guarantee synchronous dynamics between external and internal states: the spatial structure and the temporal dynamics of sensory signals and action (that comprise the system's Markov blanket) have to be pruned to produce synchrony. This synchrony is necessary for an agent to infer environmental states -- a pre-requisite for survival. Therefore, such sentient dynamics,  relies primarily on approximate synchronization between the agent and its niche.
\end{abstract}

\section{Introduction}

In any living system, many physical and chemical reactions (along with their respective transport processes) interact, leading to fluctuations and instabilities. This promotes the emergence of new traits that co-exist with other (conserved) traits. In other words, biological systems evolve due to their innate ability to sustain a dialogue with their environment \cite{Friston2009}. Interactions of a biological system with its external milieu are far from being reversible, isothermal or in equilibrium \cite{Haken1993,Nicolis1977,Prigogine1972}. These interactions entail material and energy exchange with the environment, rendering organisms thermodynamically open. 

This simple observation calls for a formal treatment of nonequilibrium steady-state (NESS) that accounts for biological self-organisation and its peculiar resistance to the second law or, more precisely, the fluctuation theorem that generalizes the second law to nonequilibrium states \cite{Seifert2012}. We address this challenge using a principle of least (variational) action or free energy in three steps: this paper represents the first step by establishing a model system that possesses the dynamical repertoire necessary to sustain nonequilibrium steady-state. Crucially, this model system has a well-defined thermodynamic free energy or Lyapunov function that we will use to characterize its evolution and convergence to nonequilibrium steady states. Our particular focus here is on equipping a system with a Markov blanket that separates internal states from external states. We illustrate the emergence of synchronization between internal and external states that is mediated through the Markov blanket; thereby modelling an elementary form of perception and action.  In brief, we will see the dynamical coupling of states separated by a Markov blanket depends on a particular sparsity and pruning of conditional dependencies that may characterise biological systems, as well as artificial systems that mimic them.

In the second step, we will treat internal states as the parameters or sufficient statistics of a probability distribution over external states, where this distribution defines variational free energy. This enables us to write the dynamics of the internal states as a variational free energy functional. We hope to show that thermodynamic and variational free energy change in the same way following external perturbations and share a common minimum at non-equilibrium steady-state. In the final step, we will show how internal states can enslave external states to minimize (variational and thermodynamic) free energy at nonequilibrium steady-state. In effect, this rests on interpreting internal states as encoding probabilistic beliefs about the external states (i.e. perception) and acting on the external states -- through the Markov blanket -- to make those beliefs come true (i.e., action). Our ultimate objective is to show that one, and only one, belief is sufficient to account for biological self-organisation; namely, the belief that the external states minimize (Shannon) entropy production.

\section{Methods}
In this section, we set up the basic form of the systems we will used to model self organisation and nonequilibrium steady-state. We start with coupled dynamics based upon the Ginzburg-Landau equation and then consider the minimal requirements for distinguishing between two sets of states; for example, states that are internal and external to the system of interest. This involves the introduction of a Markov blanket that provides a separation between internal and external states in terms of dependencies. 

\subsection{Dissipative systems -- the Ginzburg-Landau equation}

The (complex) Ginzburg-Landau equation (CGLE) describes the dynamics of a system on a manifold in a metric space that is undergoing a Hopf bifurcation from a stationary to an oscillatory state,

\begin{eqnarray}
\dot \lambda (X,t) = \lambda  + \left( {1 + i\alpha } \right){\nabla ^2}\lambda  - \left( {1 + i\beta } \right){\left| \lambda  \right|^2}\lambda 
  \label{eqn:cgle}
\end{eqnarray}

The order parameter $\lambda (X,t)$  is complex, such that  $\lambda (X,t) = \rho {e^{i\upsilon }}$. Here,  $\rho (X,t)$ and $\upsilon (X,t)$  are space and time-dependent amplitude and phase respectively. $\alpha $  and  $\beta $ measure linear and non-linear dispersion (defining the relation between frequency of the waves and the wavenumber), respectively. The real GL equation can be obtained from Eqn. \ref{eqn:cgle} by setting  $\alpha  = \beta  = 0$. The evolution of this macroscopic state can also be written in terms of a thermodynamic free energy function $\mathcal{F}$,

\begin{eqnarray}
  \dot \lambda (X,t) & = &  - \frac{{\delta \mathcal{F}}}{{\delta {\lambda ^\dag }}} \nonumber \\  
  \mathcal{F} & = & \int {{{\left| {\frac{{\partial \lambda }}{{\partial X}}} \right|}^2} - \eta {{\left| \lambda  \right|}^2} + \frac{1}{2}{{\left| \lambda  \right|}^4}} dX 
  \label{eqn:free}
\end{eqnarray}

Another limiting case of Eqn. \ref{eqn:cgle} is the conserved nonlinear Schr\"{o}dinger equation that is obtained by setting  $\alpha  = \beta  = \infty $,

\begin{eqnarray}
i\dot \lambda (X,t) = {\nabla ^2}\lambda  - {\left| \lambda  \right|^2}\lambda
  \label{eqn:schr}
\end{eqnarray}

Yet another limiting case of Eqn. \ref{eqn:cgle} is the Kuramoto-Sivashinsky (KS) equation. As the dynamics becomes unstable, any amplitude perturbation is enslaved by the phase $\varphi $  leading to the KS equation,

\begin{eqnarray}
\dot \varphi  + {\nabla ^4}\varphi  + {\nabla ^2}\varphi  + \varphi \nabla \varphi  = 0
  \label{eqn:ks}
\end{eqnarray}

Flows described by such partial differential equations (PDEs; Eqns. \ref{eqn:cgle}-\ref{eqn:ks}) occur in an infinite dimensional space. This is because if one writes sets of ordinary differential equations (ODEs) to describe such flows, an infinite set of ODEs are required to represent the dynamics of a single PDE. In practice, as dissipation gets larger, the asymptotic behaviour becomes confined to a finite dimensional inertial manifold.

\subsection{Weak (Galerkin) formulation of PDEs}

Using a Lagrange or Hermite finite-element basis, we can transform the strong form of the PDEs (Eqn. \ref{eqn:cgle}) into a variational (weak) form by multiplying by a test function on a Sobolev space  ${\mathcal{W}^1}$. We can then use the divergence theorem and discretise in time using the Crank-Nicolson method \cite{Press2007}. Analytically (and numerically), this enables one to represent Eqn. \ref{eqn:cgle} as a system of coupled systems with periodic boundaries, \textit{s.t.} $\lambda (x,y,t) = {\lambda _r}(x,y,t) + i{\lambda _c}(x,y,t)$  where   $\{ x,y\}  \subset \{ X\}$

\begin{eqnarray}
  {{\dot \lambda }_r}(x,y,t) & = & {\lambda _r} + {\nabla ^2}{\lambda _r} - {\left| \lambda  \right|^2}{\lambda _r} - \alpha {\nabla ^2}{\lambda _c} + \beta {\left| \lambda  \right|^2}{\lambda _c} \nonumber \\ 
  {{\dot \lambda }_c}(x,y,t) & = & {\lambda _c} + {\nabla ^2}{\lambda _c} - {\left| \lambda  \right|^2}{\lambda _c} + \alpha {\nabla ^2}{\lambda _r} - \beta {\left| \lambda  \right|^2}{\lambda _r}  \nonumber \\
  \label{eqn:coupl}
\end{eqnarray}

Although we describe the dynamical behaviour in ${\mathbb{C}^2}$  ($r$ denotes the real part and $c$  denotes the complex part of the field), the finite-element framework is well equipped for higher dimensions. For example, in   ${\mathbb{C}^3}$ one parcellates the domain into 6-faced elements with nodes at the vertices and use quadratic Lagrange interpolation. For simplicity, we discuss the CGLE on a 2D  domain. In our simulations, Eqn. \ref{eqn:cgle} was solved iteratively for 400 time-steps, using the generalized minimal residual method (GMRES) with incomplete LU (ILU) pre-conditioning \cite{Press2007}. Cubic-Hermite finite elements were used for increased precision. The mesh-size ($L$) was set to 512 discretisation units. Periodic boundary conditions were imposed on opposing ends.

\subsection{Positional coupling of PDEs}

Two or more PDEs (CGLE or reaction-diffusion-advection) are coupled using (positional) coupling between the real and complex valued fields as described below. The two CGLEs ($\psi$  and  $\lambda $) in Figure \ref{fig_2cgls} are reciprocally coupled as, 

\begin{strip}
\begin{eqnarray}
  \dot \psi (x,y,t)  & = & \psi  + \left( {1 + i\alpha } \right){\nabla ^2}\psi  - \left( {1 + i\beta } \right){\left| \psi  \right|^2}\psi  + d(\lambda  - \psi ) \nonumber \\ 
  \dot \lambda (x,y,t) & = & \lambda  + \left( {1 + i\alpha } \right){\nabla ^2}\lambda  - \left( {1 + i\beta } \right){\left| \lambda  \right|^2}\lambda  + d(\psi  - \lambda ) \nonumber \\
  \label{eqn:coupl2}
\end{eqnarray}
\end{strip}

Again -- for numerical simulations -- the coupled system of four PDEs (two each for the real and complex parts) was integrated for 400 time-steps, using the generalized minimal residual method (GMRES) with incomplete LU (ILU) pre-conditioning. $d$  is the coupling coefficient.

Equation \ref{eqn:coupl2} couples the modes of two systems and may thus appear sufficient to consider self organisation in terms of the coupling between a subsystem of interest (e.g.,  $\lambda$) and another; say its heat bath, econiche or external milieu (e.g.,  $\psi$). However, this setup is not sufficient to distinguish the system of interest from the system in which it is immersed. This is because there is no unique way of separating the internal states (e.g.,  $\lambda$) from external states (e.g.,  $\psi$). This follows because Eqn. \ref{eqn:coupl2} does not admit a partition of states in which internal states do not depend on external states, when conditioned upon a third set of states known as a \textit{Markov blanket}. To introduce a Markov blanket, we need to consider a partition into internal, external and Markov blanket states; where the Markov blanket states can be further partitioned into active and sensory states. Crucially, the minimal dependency structure requires that internal states cannot directly influence sensory states and external states cannot directly influence active states (see below). This ensures that the dynamics of internal states can be separated from external states but remain coupled through sensory and active states. Assigning a partition of the Markov blankets into active and sensory states appeals to the notion of an action-perception cycle; in which external states influence internal states through sensory states (i.e., perception), while internal states influence of external states through active states (i.e., action).

To create the sparse coupling necessary to support a dynamics of self organisation (i.e., the action-perception cycle in Figure \ref{fig_actpercept}) additional PDEs governing the sensory and action fields are needed. Here, we chose complex valued reaction-diffusion-advection equations of the general form, $\dot s(X,t) = f\left( {s,{\nabla _X}s,{\Delta _X}s} \right)$ and  $\dot a(X,t) = {f_2}\left( {a,{\nabla _X}a,{\Delta _X}a} \right)$. In this treatment, we use the reaction term ($s$) to represent the positional coupling. Sensation and action were represented using PDEs for two reasons: (i) in biology, both sensory receptors and motor effectors are systems that are spatially coupled in addition to their temporal dynamics, (ii) reaction-diffusion-advection underwrites most computational models in biology -- from telegraph equations (advection set to zero) describing pulse propagation in nerve cells to modelling advection and diffusion of oxygen in the vasculature.
 
Eight coupled complex-valued PDEs then provide a novel form for an (generalised) action-perception cycle; with the field  $\psi$ representing the external environment, $S$  is the sensory field,  $\lambda$ are the internal states of a system (e.g., your nervous system) and  $a$ represents action in space-time. The scalar fields ${d_{1 \ldots 4}}$  represent coupling kernels between pairs of fields, while the scalars  $\alpha $ and  $\beta $ with their respective sub-scripts represent the parameters of the CGLEs for external states and internal states.  The subscripts $r$  and $c$  denote the real and complex parts of the PDEs,

\begin{strip}
\begin{eqnarray}
  {{\dot \psi }_r} & = & {\psi _r} + {\beta _3}\left( {\psi _c^2 + \psi _r^2} \right){\psi _c} - \left( {\psi _c^2 + \psi _r^2} \right){\psi _r} - {\alpha _1}\Delta {\psi _c} + \Delta {\psi _r} + {d_1}\left( {{a_r} - {\psi _r}} \right) \nonumber \\ 
  {{\dot \psi }_c} & = & {\psi _c} - {\beta _3}\left( {\psi _c^2 + \psi _r^2} \right){\psi _r} - \left( {\psi _c^2 + \psi _r^2} \right){\psi _c} + {\alpha _1}\Delta {\psi _r} + \Delta {\psi _c} + {d_1}\left( {{a_c} - {\psi _c}} \right) \nonumber \\ 
  \nonumber \\ 
  {{\dot s}_r} & = & \Delta {s_r} + \nabla {s_r} + {d_3}\left( {{\psi _r} - {s_r}} \right) \nonumber \\ 
  {{\dot s}_c} & = & \Delta {s_c} + \nabla {s_c} + {d_3}\left( {{\psi _c} - {s_c}} \right) \nonumber \\ 
   \nonumber \\ 
  {{\dot \lambda }_r} & = & {\lambda _r} + {\beta _4}\left( {\lambda _c^2 + \lambda _r^2} \right){\lambda _c} - \left( {\lambda _c^2 + \lambda _r^2} \right){\lambda _r} - {\alpha _2}\Delta {\lambda _c} + \Delta {\lambda _r} + {d_2}\left( {{s_r} - {\lambda _r}} \right) \nonumber \\ 
  {{\dot \lambda }_c} & = & {\lambda _c} - {\beta _4}\left( {\lambda _c^2 + \lambda _r^2} \right){\lambda _r} - \left( {\lambda _c^2 + \lambda _r^2} \right){\lambda _c} + {\alpha _2}\Delta {\lambda _r} + \Delta {\lambda _c} + {d_2}\left( {{s_c} - {\lambda _c}} \right) \nonumber \\ 
  \nonumber \\
  {{\dot a}_r} & = & \Delta {a_r} + \nabla {a_r} + {d_4}\left( {{\lambda _r} - {a_r}} \right) \nonumber \\ 
  {{\dot a}_c} & = & \Delta {a_c} + \nabla {a_c} + {d_4}\left( {{\lambda _c} - {a_c}} \right) \nonumber \\ 
  \label{eqn:coupl3}
\end{eqnarray}
\end{strip}

Crucially, note that the flow of sensory states does not depend upon internal states and the flow of active states does not depend upon external states. This form therefore admits a Markov blanket and a suitable model of exchange between internal and external states. The coupling kernels $d$ allow for a spatially limited coupling with the Markov blanket that allows us to adjust both the strength and extent to which the Markov blanket states mediate between internal and external states. Having established the form of a system we will study, we now consider how to quantify coupling across the Markov blanket.

\section{Results}

In this section, we will use the Ginzburg-Landau equation to illustrate the basic dynamics afforded by this system and illustrate generalised synchrony, starting with identical synchronisation and then considering generalised synchrony. Having established the basic phenomenology, we will then look at generalised synchronisation when inserting a Markov blanket between internal and external states; i.e., in the context of self organisation through action and perception.

\subsection{The basic phenomenology of the Ginzburg-Landau equation}

Following Hohenberg and Halperin \cite{Hohenberg1977}, we study a single type of dissipative model -- Model A, representing the complex Ginzburg-Landau equation (CGLE). It is the normal form for dynamics, where fixed points lose their stability via a Hopf bifurcation, leading to a stable limit cycle oscillation. In fluid mechanics, this is known as the Newell-Whitehead equation, after the authors who derived it to study B\'{e}nard convection \cite{Newell1969}. For the CGLE, bifurcation diagrams depicting the transitions between different phases have been charted for one and two spatial dimensions \cite{Doering1994,Chate1996}. These studies have shown that for the Benjamin-Feir (BF) stable region of parameter space -- defined by $1 + \alpha  \cdot \beta  > 0$ -- one can always produce a plane wave that is linearly stable.

\begin{figure}
 \centering \includegraphics[width=0.5\textwidth, height=2in]
 {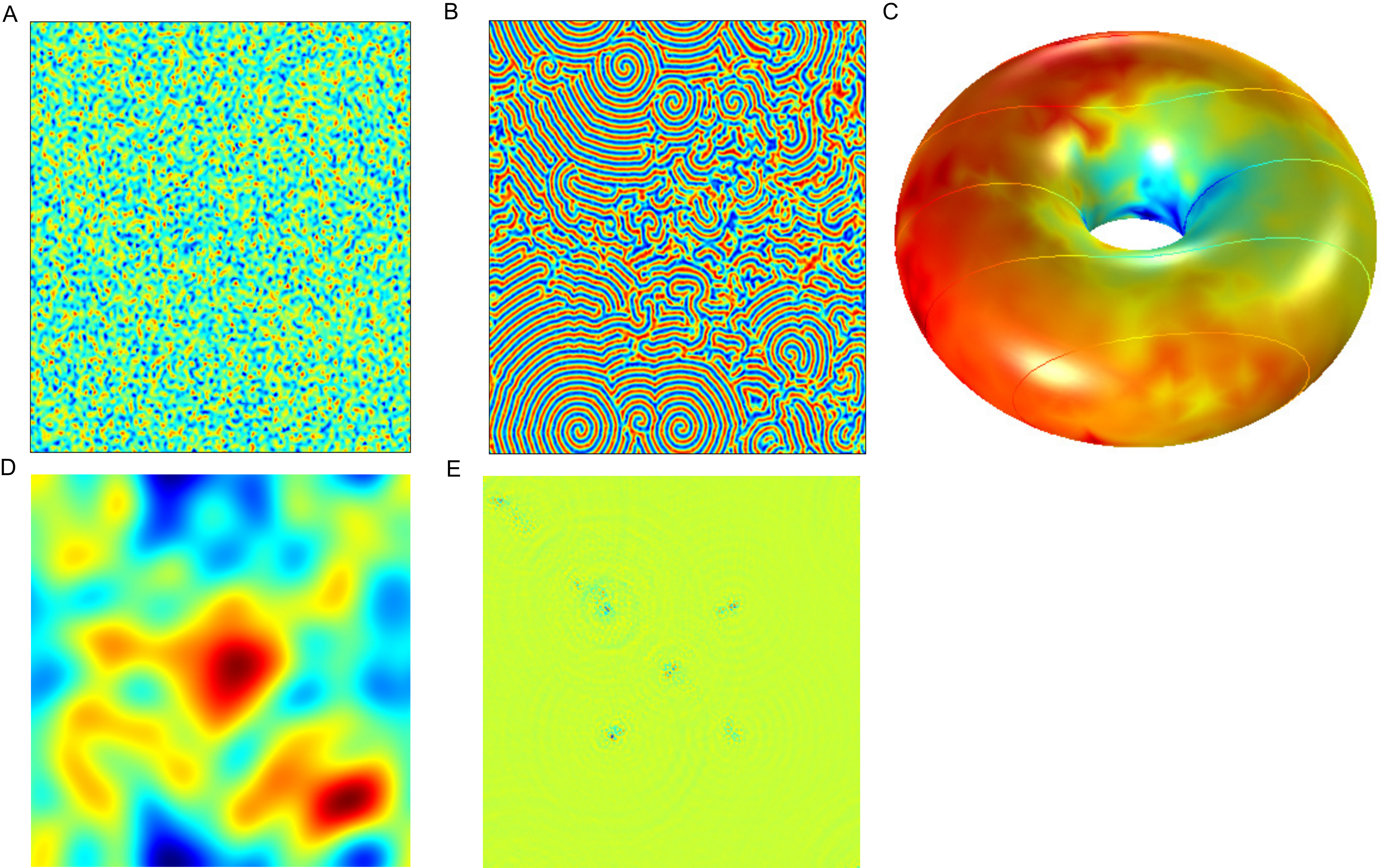}
 \caption{A 2D and a 3D (complex) Ginzburg-Landau system. In A-B and D-E the x- and y- co-ordinates of the absolute value ($\sqrt {\lambda _r^2 + \lambda _c^2} $) of the field are plotted whilst in C the z co-ordinate is plotted in addition.  Solutions at the last time-step ($T = 400$ units) are displayed for: (A) $\alpha  =  + 8,\beta  =  - 8$, (B) $\alpha  =  - 1,\beta  =  + 2$, (C) $\alpha  =  + 1,\beta  =  + 0.5$, (D) $\alpha  =  0,\beta  =  0$ (real GL) and (E) $\alpha  =  + 100,\beta  =  + 100$ (imaginary GL). Generalized minimal residual method (GMRES) with incomplete LU (ILU) pre-conditioning was used to integrate the PDEs using cubic-Hermite as the finite elements. Periodic boundary conditions were imposed with grid-size set at $512 \times 512$.}
 \label{fig_cgl}
\end{figure}

We used our simulation framework  to study the CGLE in two- and three-dimensions. As is known for the CGLE, a variety of coherent structures exist in 2D \cite{Aranson2002} depending on the values of $\alpha$ and $\beta$ in Eqn. \ref{eqn:cgle}. These coherent structures range from defect turbulence ($\left| \lambda  \right| = 0$) in Figure \ref{fig_cgl}A to spiral-waves in Figure \ref{fig_cgl}B and scroll-waves in Figure \ref{fig_cgl}C. Spiral waves are waves that rotate without changing their shape. After they destabilize in a Hopf bifurcation, they bifurcate into a drifting spiral wave.  In three-dimensions the CGLE exhibits the analogue of a 2D spiral wave in the form of a scroll wave \cite{Gabbay1997}. We simulated the CGLE on a toroid (Figure \ref{fig_cgl}C) and observed turbulence of scroll rings around the toroid (Figure \ref{fig_cgl}C). Unlike spiral waves, scroll waves have more degrees of freedom -- enabling them to flow in space with varying shapes. As noted by Aranson and Kramer \cite{Aranson2002}, the 3D-CGLE also displays vortices that can be highly unstable for the range of parameters over which their 2D analogue is completely stable. Setting, $\alpha = 0$, $\beta = 0$ gives us the real Ginzburg-Landau equation (Figure \ref{fig_cgl}D), representative of coherent brain states of the sort measured in electrophysiology. Similarly, as $\alpha$ and $\beta$ attain larger values, the CGLE equation becomes conservative (i.e., Hamiltonian) and we can approximate the eponymous nonlinear Schr\"{o}dinger equation (Figure \ref{fig_cgl}E). 

In summary, this single equation is capable of exhibiting a wide variety of spatiotemporal structures (\textit{cf.} deep recurrent networks); including deterministic chaos: this is the primary reason that we have selected the CGLE to represent an interacting agent and its environment. Furthermore, this equation can be reduced to a neural field equation (private communication, Jack Cowan) that allows us to place what follows in a neural context; in other words, it can represent millions of neurons interacting with each other. Our intention here is not to meticulously describe the bifurcation structure of this equation but to use it as a model system to study the self-organisation of systems and their approximate milieu that display similar bifurcations and -- most importantly -- are dissipative.

\subsection{Coupling and synchrony in spatially extended systems}

\begin{figure}
 \centering \includegraphics[width=0.5\textwidth, height=1.5in]
 {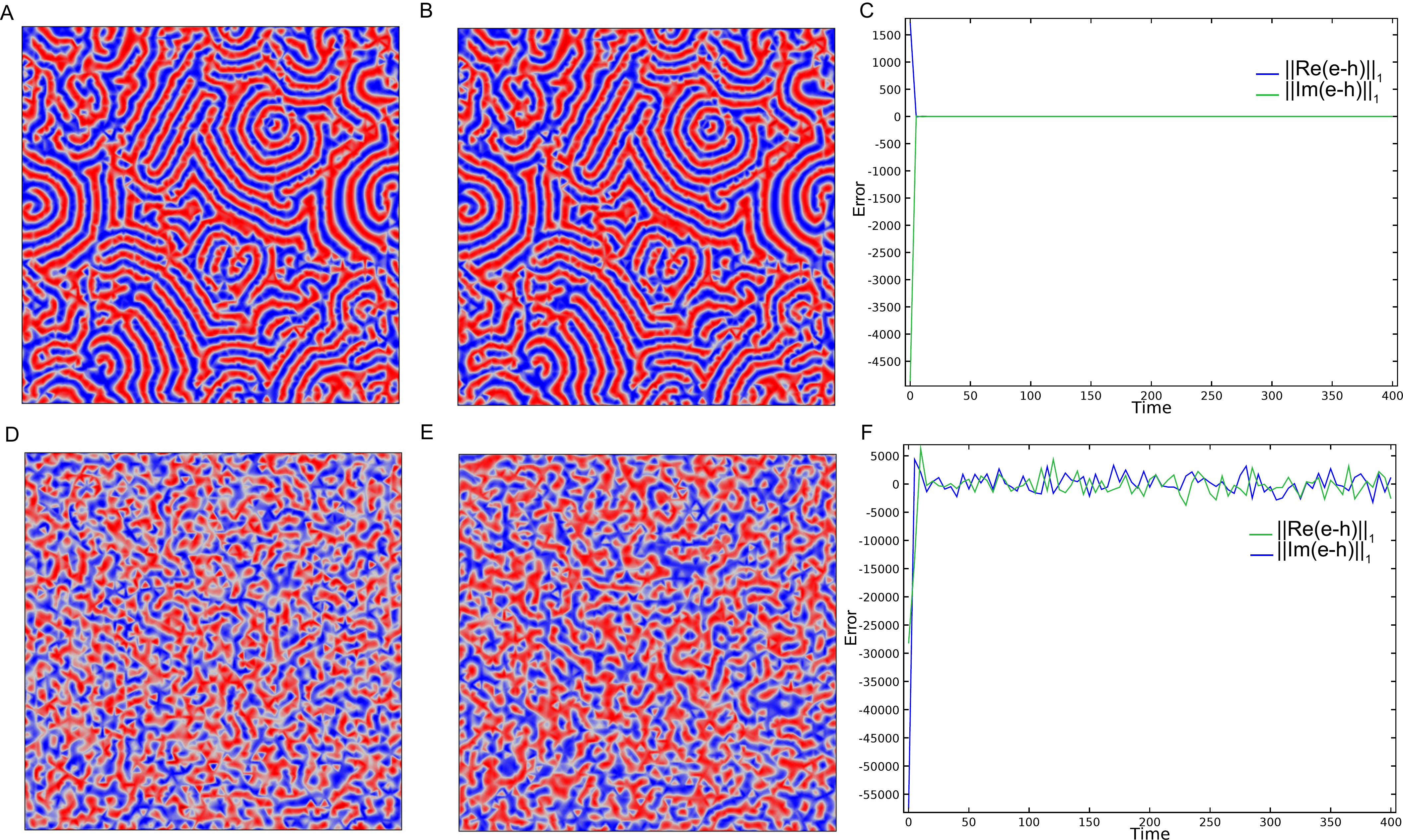}
 \caption{Two 2D CGLE reciprocally coupled using positional interactions i.e., $d({\lambda _{2,1}} - {\lambda _{1,2}})$ with $d=1$, where ${\lambda _1}$ and ${\lambda _2}$ describe the individual CGLEs. The x- and y- axis represent the absolute value of the field.  (A,B) Two CGLE with identical parameters -- $\alpha = -1$,$\beta = +2$. (C) The difference between the real and the imaginary parts of fields shown in A and B. (D,E) Two CGLE with non-identical parameters -- (D) has $\alpha = +2$,$\beta = -6$ and (E) has $\alpha = -1$,$\beta = -9$. (F) Same as C but between non-identical fields.}
 \label{fig_2cgls}
\end{figure}

To simulate self-organisation, it is necessary to distinguish between the system \textit{per se} and its (proximate) environment. Thermodynamically speaking, the environment acts as a heath bath for the agent wherein the agent accrues energy for its survival from the environment. In order to exchange information it is vital that the spatiotemporal dynamics of both parties are coordinated; i.e., they are synchronized. We will start with the simplest form of generalised synchrony, known as identical synchronization because the states or fields of the system and its environment become identical over time. In other words, the motion in the joint space of the system and its environment are constrained to an attracting set on a hyperplane, known as the synchronization manifold. When the environment and the agent have identical chaotic spatiotemporal dynamics (i.e., the parameters of the equations above are the same), it takes less than 20 time-units of simulation for two reciprocally coupled CGLE to reduce real and complex errors to zero. This is shown in Figure \ref{fig_2cgls}A-C, where the synchronisation error corresponds to the difference between the states or distance from the identity synchronisation manifold (as quantified by the surface integral of the difference between the respective real and the imaginary parts of the equation). 

More often than not, coupled systems have different parameters; i.e., they have different spatiotemporal dynamics. Generally these systems develop some sort of (generalised) synchrony as time unfolds. Indeed, in two different but reciprocally coupled CGLEs, both the real and the imaginary parts approach zero with time, yet they are never identical. This means there exists a smooth invertible mapping  $\Phi :\psi  \to \lambda$ between the fields representing the environment ($\psi$) and the system or agent ($\lambda$). Such a form of synchrony is known as generalized synchronization \cite{Hunt1997,Barreto2003}. This is useful, as it tells us that the agent can predict the dynamics of the environment: for every point on $\psi$ there is another point on $\lambda$ that lies on the synchronization manifold.

\subsection{The Markov blanket}

\begin{figure}
 \centering \includegraphics[width=0.5\textwidth, height=2in]
 {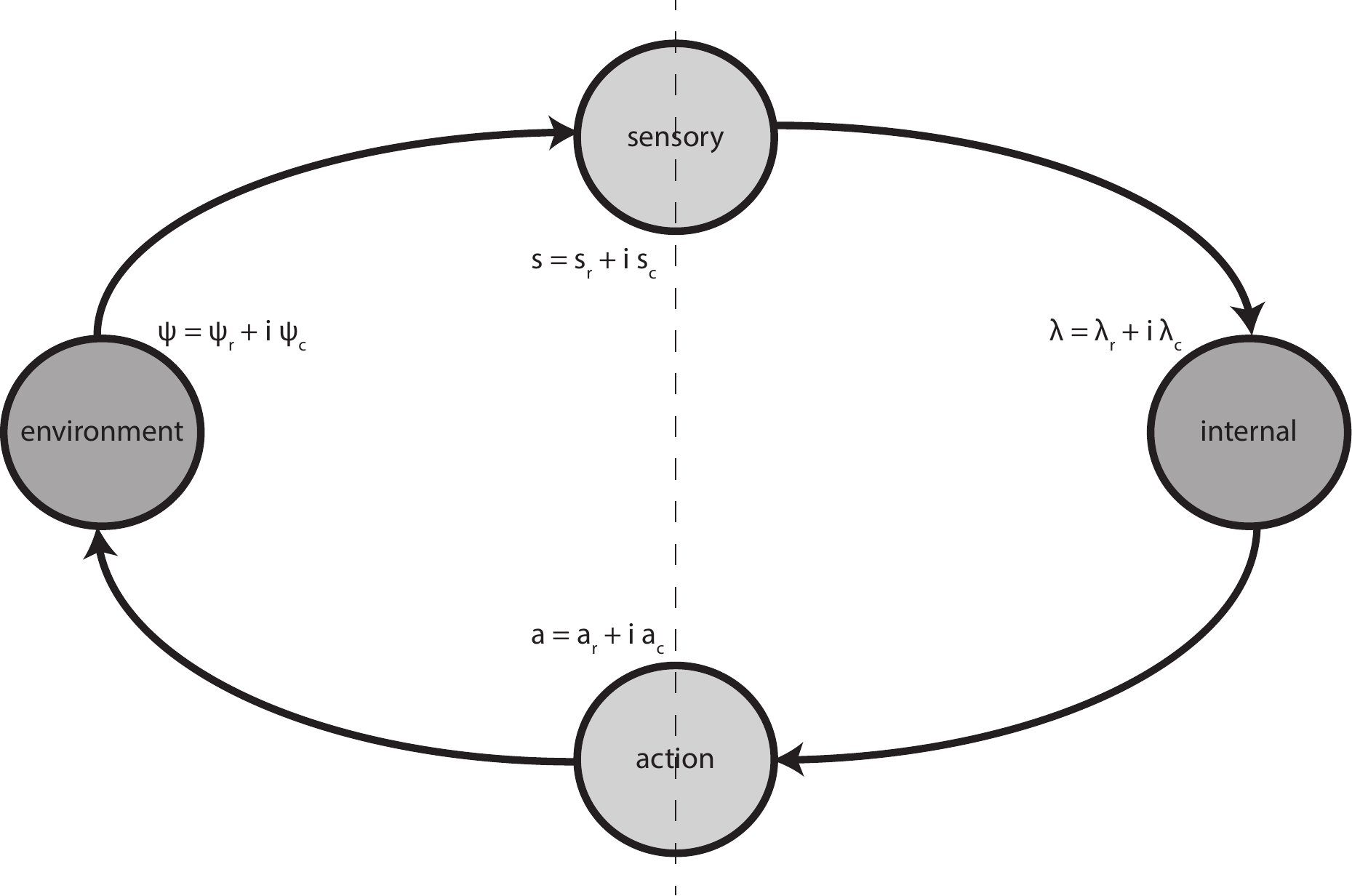}
 \caption{The Action-Perception cycle. Two complex-valued Ginzburg-Landau equations representing the environment and the agent's internal states are coupled via spatiotemporal states that constitute its Markov blanket; i.e., the sensory and action fields. Here, we represent sensation and action as arbitrary complex-valued reaction-diffusion-advection equations, to model the spatiotemporal dynamics of sensation and action. For details please see Eqn.   \ref{eqn:coupl3}.}
 \label{fig_actpercept}
\end{figure}

Statistically, a reciprocally coupled dynamical system requires the existence of a Markov blanket to emulate self-organization (Figure \ref{fig_actpercept}) \cite{Sengupta2016}. In other words, to distinguish internal (agent) and external (environmental) states statistically, we have to consider the Markov blanket that separates them. The Markov blanket is defined in terms of dependencies that can be modelled in terms of the equations of motion or flow underlying the dynamics in question. The Markov blanket of any subset of states is defined as its children, its parents and the children of the parents  \cite{Pearl1988}. There are many ways in which one could model a partition between internal and external states. Here, we elected to do this by coupling two non-identical systems of the sort described in the previous section through additional states that have the necessary dependencies to form a Markov blanket. The systems described in Figure \ref{fig_2cgls} do not possess a Markov blanket i.e., they complete knowledge of one another and cannot be separated in a meaningful way. To introduce a Markov blanket we turn to Eqn.   \ref{eqn:coupl3}, where the Markov blanket is partitioned into states that are not influenced by internal states (sensory states) and states that are not influenced by external states (active states). Heuristically, this means that the internal states (e.g., agent) and external states (e.g., its environment) have vicarious knowledge of one another -- it is only through the sensory states (e.g., sensors for sensory receptors or cameras/LIDAR for artificial agents) that the agent acquires knowledge of its environment. It uses this partially observed knowledge to elaborate an action, mediated through active states (e.g., motor effectors for biological agents or controllers for artificial agents). Note that this partition is symmetrical with respect to the dependency structure -- the fundamental distinction between internal and external states will become apparent in subsequent papers. Note also that the presence of generalised synchrony implies the existence of an attracting set and therefore weakly mixing ergodicity. This means that if generalised synchrony can be maintained through a Markov blanket, the system self organises in a fundamental (ergodic) sense through the existence of an attracting set. This is why we consider generalised synchrony to be such an important characteristic of self organisation.

A crucial aspect of considering sensory and active components of the Markov blanket is that the internal and external states are reciprocally coupled. In other words, we do not consider a skew-product (master-slave) system -- but two systems that influence and are influenced by each other (circular causality). Our interest here is in the way that they synchronise and how this synchronisation reflects the emergence of a non-equilibrium steady state. This nonequilibrium steady-state is synonymous with the attracting set or synchronisation manifold that underwrites generalised synchrony. The reciprocal coupling across the Markov blanket means that there is a circular causality in play. In other words, in terms of the action perception cycle, nonequilibrium steady-state rests on both the agent (i.e., internal states) acting on the environment (i.e., external states) and the environment acting on the agent.

For the CGLE model, two non-identical systems (on two-dimensional manifolds) were coupled using auxiliary states that constitute the Markov blanket as follows -- complex-valued sensory and action states were introduced to play the role of sensation (via sensory receptors) and motor action (via effector organs), respectively. The sensory state is governed by arbitrary reaction-diffusion-advection dynamics (see Methods) but depends on the agent's environment, while active states depend on the internal states of the agent. Both sensory and the action fields are endowed with a spatial structure due to the fact that sensory receptors and motor effectors have spatiotemporal (and not just temporal) dynamics. Let us motivate this construction  -- most creatures do not have a single photo-receptor to detect light but an array of receptors that have a unique spatial and temporal architecture \cite{Land2012}. Similarly, motor effectors such as joints require co-ordination in space as well as time to enable locomotion \cite{Alexander1996}. Such a coupled system then forms an explicit way of perturbing the environment based on the information accrued by the agent. 

\begin{figure}
 \centering \includegraphics[width=0.5\textwidth, height=2in]
 {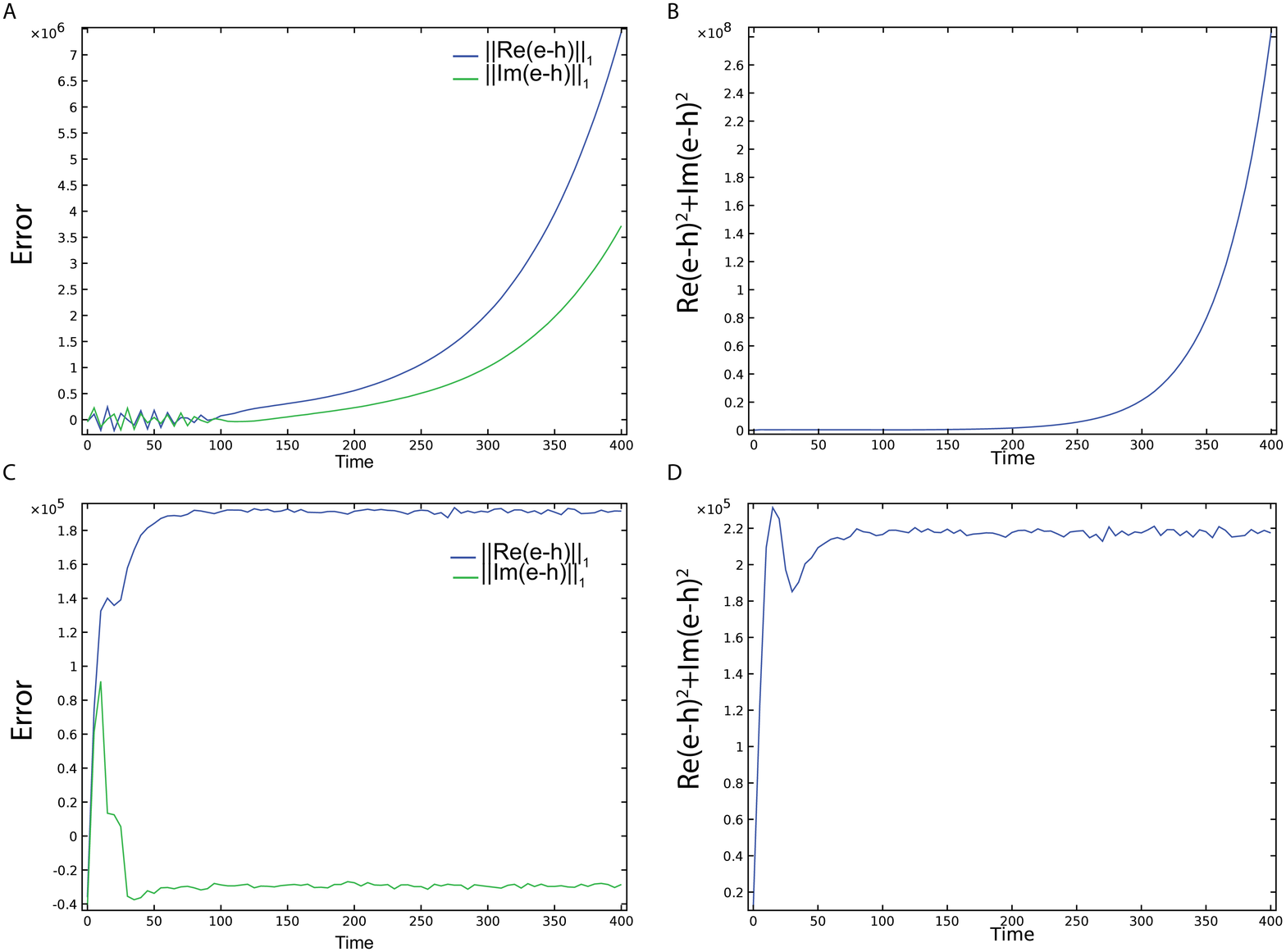}
 \caption{(A) The error between the real and imaginary parts of the environment and the internal states. (B) The error magnitude between the fields in (A). (C) Same as A -- but with the coupling coefficient set at $d=0.3$ instead of $d=0.003$. (D) Same as B with $d=0.3$ coupling coefficient.}
 \label{fig_norms}
\end{figure}

Coupling the CGLE equations through a Markov blanket (represented as reaction-diffusion equations as per Figure \ref{fig_actpercept}) with low coupling strength results in an exponential divergence between the external and the internal states (Figure \ref{fig_norms}A-B). This is generally expected when deterministic chaos in one field is unable to attract the dynamics of the reciprocally connected field onto the synchronization manifold. This is because the phase space keeps expanding. In this scenario, no synchronisation between the environment and the agent is seen (Figure \ref{fig_norms}A-B). Trivially, increasing the coupling coefficient from 0.003 to 0.3 remedies the exponential divergence; such that error between the real and the imaginary parts of the internal and external fields remain bounded (Figure \ref{fig_norms}C) and, most importantly, constant (Figure \ref{fig_norms}D).

In fact, one can obtain identical synchronization by altering the dynamics of the sensory and the action fields, thereby changing how they (the Markov blanket) are influenced by the environment and the internal states. Specifically, we replaced the sensory (equivalently for action) fields in Eqn. \ref{eqn:coupl3}  by

\begin{eqnarray}
  {{\dot s}_r} & = & \Delta {s_r} + \nabla {s_r} + {d_3}{\psi _r} \nonumber\\ 
  {{\dot s}_c} & = & {d_3}{\psi _c} 
  \label{eqn:diffcoupling}
\end{eqnarray}

\begin{figure}
 \centering \includegraphics[width=0.5\textwidth, height=2in]
 {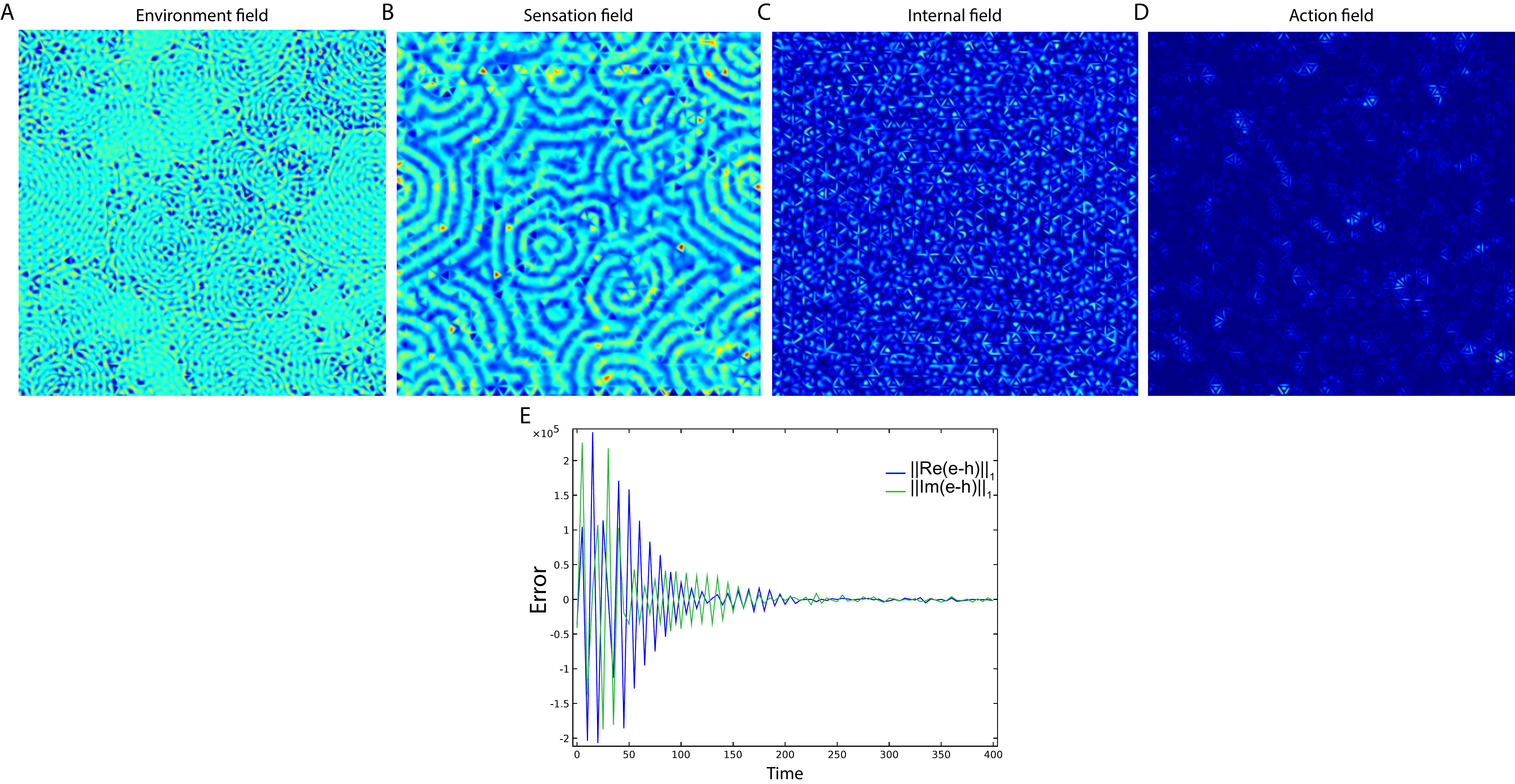}
 \caption{Coupled CGLE as per the scheme in Figure \ref{fig_actpercept} but with Eqn. \ref{eqn:coupl3} replaced by Eqn. \ref{eqn:diffcoupling} i.e., when the dynamics of the Markov blanket is simplified to suppress their memory. A-D represent the external, sensory, internal and action fields respectively. Coupling between fields was set to $d=0.003$. (E) The error  between the real and imaginary parts of the external and the internal fields.}
 \label{fig_4cgls}
\end{figure}

This effectively suppresses the decay of the sensory and active states and simplifies the spatial memory of their imaginary part. In effect, this renders sensory and active states a more direct reflection of external and internal states respectively.

Using such a dynamics for the sensory and action fields mean that the divergence between the environment and internal state fields tends to zero over time (Figure \ref{fig_4cgls}) -- a characteristic of a specific form of generalized synchrony (identity mapping) under weak coupling. In other words, generalised synchrony rests upon particular conditions on the coupling under this generic model of self organisation.

This concludes our illustration of non-equilibrium steady-state dynamics in dynamical systems that are coupled through a Markov blanket. 

\section{Discussion}

The complex-valued Ginzburg-Landau equation (CGLE) offers a numerical framework to pose and study the interaction of an agent with its environment (for example, to analyze reinforcement learning systems). Quite naturally, both parties display rich behaviour in the form of phase transitions (bifurcations). These phase transitions range from spiral wave formation in two dimensions to the inception of scroll waves in three dimensions. We have shown how coupling of two dissipative systems lead to particular forms of synchronization. However, a direct mapping between the external states of the environment and the internal states of the agent renders them formally inseparable; requiring us to introduce the notion of Markov blankets. Here, we have used two intermediary reaction-diffusion-advection equations; namely, sensory and action fields. Such a setup supports a generalised action-perception cycle in a space-time dependent formalism. Our numerical simulations of biological self-organization via the dynamics of the action (motor) -- perception (sensory) fields highlight the fact that synchronization between the environment and an agent can only occur with specific forms of coupling, through the Markov blanket. This is important because it suggests the emergence of self-organisation is non-trivial; requiring the evolution of particular mappings (receptors and appendages) that guarantee synchronization between the agent and its ecological niche. In the absence of synchronization, the agent is simply unequipped to infer and learn about its local niche and therefore cannot adapt itself to sustain itself under burgeoning uncertainties.

By altering two parameters ($\alpha$ and $\beta$) the CGLE equation can exhibit phase turbulence, defect turbulence, spatiotemporal intermittency and bichaos, many of which are common in systems operating out of equilibrium \cite{Sakaguchi1990,Shraiman1992,Egolf1995}. These recurring instabilities require a continuous supply of matter or energy. They cease to exist as soon as the supply of energy is removed. For infinite dimensional systems, such as PDEs, and invariably for biological organisms, self-organization can therefore be understood as the convergence of the dynamics under certain initial conditions to solutions that are stable and synchronized under external perturbations.

Our numerical experiments speak to the hypothesis that predictability is intertwined with generalized synchronization; i.e., there can be no reduction in uncertainty about the environment encoded by the internal states of the agent unless the dynamics of both the agent and environment have settled onto a synchronization manifold. The intuition follows from the fact that in coupled chaotic systems the internal state-space can be exponentially expanding; any inferential mechanism that reduces uncertainty works on the premise that the system being inferred has certain properties (say being smooth, continuous, existence of Markov partition, etc.). This means by constantly changing sensations and generating actions the internal states of the agent ought to squeeze the internal states to a synchronization manifold. Squeezing is used in the study of dissipative dynamics to portray the evolution of a trajectory to an inertial manifold. In short, our working hypothesis is inference (instantiated as some form of energy minimization) enables the agent to squeeze dissipative states into a synchronization manifold. We will show in forthcoming papers that it is the existence of such a manifold that enables one to minimize the prediction error (using MCMC or variational Bayes) in order to prescribe the most probable internal states that mimic the environment.

The construction that we have adopted to study self-organization assumes that the spatial extent of the sensory field is equivalent to the environmental field -- or a restricted version of it; i.e., they are mapped to one another in a one-to-one fashion. It would be an interesting exercise to study how different spatial forms (kernels) of coupling between the environment and the sensations or between the internal states of the agent and the action give rise to a zoo of spatiotemporal pattern formation. The practical motivation behind the simple coupling kernels considered in this paper is numerical -- using identically sized finite-element mesh with simple coupling bounds the stiffness of the underlying differential equations, enabling us to iteratively solve the coupled system of eight partial differential equations (i.e., action-perception cycle) in a reasonable amount of time. However, our sensory receptors only have access to restricted and nonlinear samples from the environment (a bottleneck) and it will be important to study the effect of punctuated coupling kernels. 

We may have modelled the action-perception cycle in purely phenomenological terms yet the principles manifest at this level percolate to lower levels (under a change of gauge; see \cite{Sengupta2016}). For example, chemical reactions in biological systems do not take place homogeneously over a surface but via propagation of waves of oxidation that travel across the reaction surface. The non-linearity in chemical kinetics, the diffusion of chemical species -- and lastly, the external perturbations that force the reaction to occur in far-from-equilibrium conditions, are the major ingredients for biotic self-organization. Equally, it is important to realize that it is not a single instability but a hierarchy of instabilities that enables a biological system to spontaneously produce increasingly complex modes. This happens as the biological organism dissipates more and more energy -- after instability -- creating conditions favourable for more instability.

It has been  argued that the thermodynamic (Helmholtz) free-energy and variational free-energy share the same minimum at thermodynamic equilibrium \cite{Sengupta2013}. Yet a direct connection between non-equilibrium entropy production and Shannon entropy has been missing in the context of biological self-organization and evolution. Our formalism allows one to develop intuitions about how the two quantities could be linked. Given that a thermodynamic free-energy functional can be written for the complex valued Ginzburg-Landau equation \cite{Descalzi1992} and variational inference methodology can be applied to formulate a free-energy functional over the environmental states; both could be compared qualitatively using analysis as well as numerical simulations. This puts us in a unique position to compare thermodynamic and informational energies for the first time.

It is well-known that under the chaotic hypothesis (\cite{Gallavotti1995}; for more general fluctuation theorems see \cite{Sinai1968,Jarzynski1997,Crooks1999,Evans2002,Seifert2012}), the evolution of internal states will invariably converge to a non-equilibrium steady state (one of them being the Sinai-Ruelle-Bowen (SRB) measure \cite{Sinai1968,Bowen2004}). Biological systems and artificial decision making systems are no different. Crucially, under Hamiltonian Anosov flows, the chaotic hypothesis reduces to the ergodic hypothesis of equilibrium thermodynamics \cite{Birkhoff1931} and converges to a Boltzmann-Gibbs measure. In our action-perception cycle (Figure \ref{fig_actpercept}) the situation is a little more involved, as the PDEs describing the internal states indirectly (via the sensory forcing) embed the dynamics of the reservoir (the action field). We argue that sensations drive the internal states of the agent to a non-equilibrium steady state (with a valid SRB measure). This enables the internal states to cool the information received from the sensorium wherein the extra heat (entropy) produced is returned as action, the thermal reservoir in this microcosm. 

In summary, we have taken the first step in formulating a framework where the interplay between thermodynamic and variational energies could be studied. The second step is to compare and contrast these energies numerically and analytically under a variety of parameter combinations and coupling kernels, esp. those kernels that impose a bottleneck. 

\bibliographystyle{ieee}
\bibliography{sentient_intelligence}

\begin{thebibliography}{10}\itemsep=-1pt

\bibitem{Alexander1996}
R.~M. Alexander.
\newblock {\em Optima for Animals}.
\newblock Princeton University Press, 1996.

\bibitem{Aranson2002}
I.~S. Aranson and L.~Kramer.
\newblock The world of the complex {Ginzburg-Landau} equation.
\newblock {\em Rev. Mod. Phys.}, 74:99--143, 2002.

\bibitem{Barreto2003}
E.~Barreto, K.~Josić, C.~J. Morales, E.~Sander, and P.~So.
\newblock The geometry of chaos synchronization.
\newblock {\em Chaos}, 13(1):151--164, 2003.

\bibitem{Birkhoff1931}
G.~D. Birkhoff.
\newblock Proof of the ergodic theorem.
\newblock {\em Proceedings of the National Academy of Sciences},
  17(12):656--660, 1931.

\bibitem{Bowen2004}
R.~Bowen and D.~Ruelle.
\newblock {\em The Ergodic Theory of Axiom A Flows}, book section~5, pages
  55--76.
\newblock Springer New York, 2004.

\bibitem{Chate1996}
H.~Chate and P.~Manneville.
\newblock Phase diagram of the two-dimensional complex {Ginzburg-Landau}
  equation.
\newblock {\em Physica A}, 224:348--368, 1996.

\bibitem{Crooks1999}
G.~E. Crooks.
\newblock Entropy production fluctuation theorem and the nonequilibrium work
  relation for free energy differences.
\newblock {\em Phys Rev E}, 60(3):2721--2726, 1999.

\bibitem{Descalzi1992}
O.~Descalzi and R.~Graham.
\newblock Gradient expansion of the nonequilibrium potential for the
  supercritical {Ginzburg-Landau} equation.
\newblock {\em Physics Letters A}, 170(2):84--90, 1992.

\bibitem{Doering1994}
C.~R. Doering, J.~D. Gibbon, and C.~David~Levermore.
\newblock Weak and strong solutions of the complex {Ginzburg-Landau} equation.
\newblock {\em Physica D}, 71:285--318, 1994.

\bibitem{Egolf1995}
D.~A. Egolf and H.~S. Greenside.
\newblock Characterization of the transition from defect to phase turbulence.
\newblock {\em Phys Rev Lett}, 74(10):1751--1754, 1995.

\bibitem{Evans2002}
D.~J. Evans and D.~J. Searles.
\newblock The fluctuation theorem.
\newblock {\em Advances in Physics}, 51(7):1529--1585, 2002.

\bibitem{Friston2009}
K.~Friston.
\newblock The free-energy principle: a rough guide to the brain?
\newblock {\em Trends Cogn Sci}, 13(7):293--301, 2009.

\bibitem{Gabbay1997}
M.~Gabbay, E.~Ott, and P.~N. Guzdar.
\newblock Motion of scroll wave filaments in the complex {Ginzburg-Landau}
  equation.
\newblock {\em Physical Review Letters}, 78(10):2012--2015, 1997.

\bibitem{Gallavotti1995}
G.~Gallavotti and E.~G.~D. Cohen.
\newblock Dynamical ensembles in nonequilibrium statistical mechanics.
\newblock {\em Phys Rev Lett}, 74(14):2694--2697, 1995.

\bibitem{Haken1993}
H.~Haken.
\newblock {\em Advanced Synergetics: Instability Hierarchies of Self-Organizing
  Systems and Devices}.
\newblock Springer-Verlag, New York, 1993.

\bibitem{Hohenberg1977}
P.~C. Hohenberg and B.~I. Halperin.
\newblock Theory of dynamic critical phenomena.
\newblock {\em Rev Mod Phys}, 49:435--479, 1977.

\bibitem{Hunt1997}
B.~R. Hunt, E.~Ott, and J.~A. Yorke.
\newblock Differentiable generalized synchronization of chaos.
\newblock {\em Physical Review E}, 55(4):4029--4034, 1997.

\bibitem{Jarzynski1997}
C.~Jarzynski.
\newblock Nonequilibrium equality for free energy differences.
\newblock {\em Phys Rev Lett}, 78(14):2690--2693, 1997.

\bibitem{Land2012}
M.~F. Land and D.-E. Nilsson.
\newblock {\em Animal Eyes}.
\newblock Oxford University Press, 2012.

\bibitem{Newell1969}
A.~Newell and J.~Whitehead.
\newblock Finite bandwidth, finite amplitude convection.
\newblock {\em Journal of Fluid Mechanics}, 38(2):279--303, 1969.

\bibitem{Nicolis1977}
G.~Nicolis and I.~Prigogine.
\newblock {\em Self-Organization in Non-Equilibrium Systems}.
\newblock Wiley, New York, 1977.

\bibitem{Pearl1988}
J.~Pearl.
\newblock {\em Probabilistic reasoning in intelligent systems: networks of
  plausible inference}.
\newblock Morgan Kaufmann Publishers Inc., 1988.

\bibitem{Press2007}
W.~H. Press, S.~A. Teukolsky, W.~T. Vetterling, and B.~P. Flannery.
\newblock {\em Numerical Recipes: The Art of Scientific Computing}, volume~3.
\newblock Cambridge University Press, New York, NY, USA, 3 edition, 2007.

\bibitem{Prigogine1972}
L.~Prigogine, G.~Nicolis, and A.~Babloyantz.
\newblock Thermodynamics of evolution.
\newblock {\em Physics Today}, 25(12):38--44, 1972.

\bibitem{Sakaguchi1990}
H.~Sakaguchi.
\newblock Breakdown of the phase dynamics.
\newblock {\em Prog Theor Phys}, 84:792--800, 1990.

\bibitem{Seifert2012}
U.~Seifert.
\newblock Stochastic thermodynamics, fluctuation theorems and molecular
  machines.
\newblock {\em Rep Prog Phys}, 75(12):126001, 2012.

\bibitem{Sengupta2013}
B.~Sengupta, M.~B. Stemmler, and K.~J. Friston.
\newblock Information and efficiency in the nervous system–a synthesis.
\newblock {\em PLoS Comput Biol}, 9(7):e1003157, 2013.

\bibitem{Sengupta2016}
B.~Sengupta, A.~Tozzi, G.~K. Cooray, P.~K. Douglas, and K.~J. Friston.
\newblock Towards a neuronal gauge theory.
\newblock {\em PLoS Biology}, 14(3):1--12, 03 2016.

\bibitem{Shraiman1992}
B.~I. Shraiman, A.~Pumir, W.~van Saarloos, P.~C. Hohenberg, H.~Chaté, and
  M.~Holen.
\newblock Spatiotemporal chaos in the one-dimensional complex ginzburg-landau
  equation.
\newblock {\em Physica D}, 57:241--248, 1992.

\bibitem{Sinai1968}
Y.~G. Sinai.
\newblock Markov partitions and c-diffeomorphisms.
\newblock {\em Functional Analysis and Its Applications}, 2(1):61--82, 1968.

\end{thebibliography}

\end{document}